\documentclass[twocolumn,aps, prl, superscriptaddress, showpacs, amsmath,amssymb]{revtex4}
\usepackage{graphicx}
\usepackage{dcolumn}
\usepackage{bm}
\begin{document}
\title{Three-dimensional shear in granular flow}
\author{Xiang Cheng}
\affiliation{The James Franck
Institute and Department of Physics, The University of Chicago,
Chicago, IL 60637}
\author{Jeremy B. Lechman}
\affiliation{Sandia National
Laboratories, Albuquerque, NM 87185}
\author{Antonio F. Barbero}
\altaffiliation [Permanent address: ] {Complex Fluids Physics
Group, Department of Applied Physics, University of Almeria,
Almeria 04120, Spain.} \affiliation{The James Franck Institute and
Department of Physics, The University of Chicago, Chicago, IL
60637}
\author{Gary S. Grest}
\affiliation{Sandia National Laboratories, Albuquerque, NM 87185}
\author{Heinrich M. Jaeger}
\affiliation{The James Franck Institute and Department of Physics,
The University of Chicago, Chicago, IL 60637}
\author{Greg S. Karczmar}
\affiliation{Department of Radiology, The University of Chicago,
Chicago, IL 60637}
\author{Matthias E. M\"{o}bius}
\affiliation{The James Franck Institute and Department of Physics,
The University of Chicago, Chicago, IL 60637}
\author{Sidney R. Nagel}
\affiliation{The James Franck Institute and Department of Physics,
The University of Chicago, Chicago, IL 60637} \pacs{45.70.Mg,
45.70.-n, 83.50.Ax}
\date{\today}

\begin{abstract}
The evolution of granular shear flow is investigated as a function
of height in a split-bottom Couette cell. Using particle tracking,
magnetic-resonance imaging, and large-scale simulations we find a
transition in the nature of the shear as a characteristic height
$H^*$ is exceeded. Below $H^*$ there is a central stationary core;
above $H^*$ we observe the onset of additional axial shear
associated with torsional failure. Radial and axial shear profiles
are qualitatively different: the radial extent is wide and
increases with height while the axial width remains narrow and
fixed.
\end{abstract}

\maketitle Shear bands in dense granular materials are localized
regions of large velocity gradients; they are the antithesis of
the broad uniform flows seen in slowly-sheared Newtonian fluids
\cite{14,1,2,3,4,5}.  Until recently it was generally assumed that
all granular shear bands were narrow.  However, in 2003 Fenistein
{\it et al.}\cite{6} discovered that in modified Couette cells
granular shear bands can be made arbitrarily broad. In this
geometry, the bottom of a cylindrical container is split at radius
$r=R_s$ and shear is produced by rotating both the outer ring and
the cylindrical boundary of the container while keeping the
central disk ($r < R_s$) stationary. For very shallow packs, the
shear band measured at the top surface is narrow and located at $r
= R_s$ so that the inner region directly above the central disk is
stationary while the remaining part rotates as a solid. As the
filling height of the material, $H$, increases, the shear band
increases in radial width and moves toward the cylinder axis. For
sufficiently large $H$, the shear band overlaps the axis at $r=0$
and one might expect qualitatively new behavior. Indeed, Unger
{\it et al.}\cite{7} predicted that the shape of the boundary
between moving and stationary material would undergo a first-order
transition as $H$ is increased past a threshold value $H^*$: the
shearing region which for $H < H^*$ is open at the top and
intersects the free surface abruptly collapses to a closed cupola
completely buried inside the bulk.

Previous experiments focused primarily on the surface flows in
shallow containers and left unexplored many questions about the
shape and evolution of the shear profiles for large $H$. Here, we
combine magnetic resonance imaging (MRI) and high-speed video
observations with large-scale simulations to explore shear flow
both for shallow and tall packs.  In addition to monitoring the
evolution of the flow profiles in the radial direction, we also
examine shear in the vertical direction.  Instead of a first order
collapse of the shear zone as proposed by Unger {\it et
al.}\cite{7}, we find that above $H^* \simeq 0.6R_s$, the inner
core of immobile material disappears gradually as shear along the
central axis of the cylinder sets in.

\begin{figure}[t!]
\begin{center}
\includegraphics[width=2.9in]{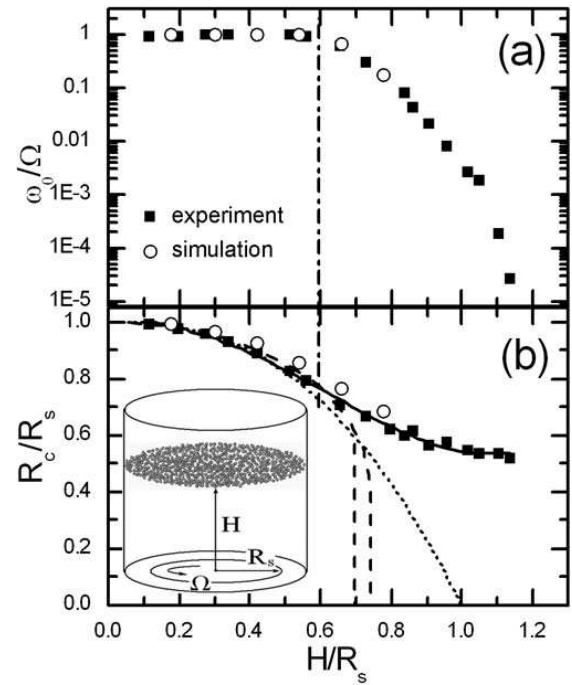}%
\caption{\label{setup} Surface flow as function of filling height,
$H$. (a) Angular velocity, $\omega_0$ at the cell axis. (b) Center
of the shear band, $R_c$. Solid line is a gaussian fit to guide
the eye; dotted line is the fitting function of Fenistein {\it et
al.} \cite{6}; dashed line is the theoretical result \cite{7}.
Vertical dash-dot line indicates $H=0.6R_s$. Inset: Schematic of
split-bottom Couette cell.}
\end{center}
\end{figure}

\begin{figure}
\begin{center}
\includegraphics[width=2.9in]{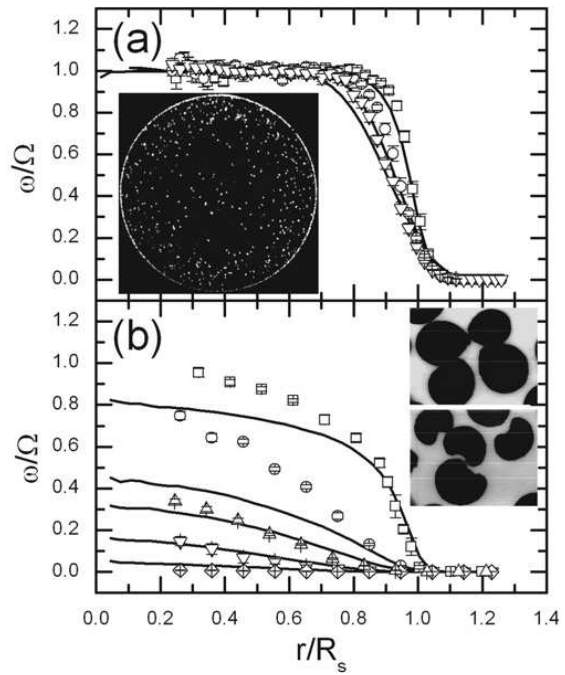}%
\caption{\label{figure2} Angular velocity profiles $\omega(r)$ at
different vertical positions $h$ for (a) $H=0.4R_s$ and (b)
$H=0.88R_s$. Symbols for MRI experiments: (a) $h=0.21H$
($\square$), $h=0.43H$ ($\circ$) and $h=0.65H$ ($\triangledown$);
(b) $h=0.10H$ ($\square$), $h=0.21H$ ($\circ$), $h=0.31H$
($\triangle$), $h=0.42H$ ($\triangledown$) and $h=0.73H$
($\diamond$). Lines for simulations: (a) from right to left
$h=0.19H$, $h=0.50H$ and $h=0.85H$; (b) from top to bottom
$h=0.11H$, $h=0.23H$, $h=0.30H$, $h=0.43H$ and $h=0.70H$. (a)
Inset: MRI image for one layer inside the bulk. Bright spots are
poppy seeds, dark background are rajagara seeds; poppy seeds glued
to the wall of the cell show up as a bright circle. (b) Inset:
Optical micrographs of rajagara (top) and poppy seeds (bottom).}
\end{center}
\end{figure}

Our setup is similar to that of Fenistein {\it et al.}\cite{6}
except that we rotate the inner disk instead of the outer ring and
cylinder (Fig.\ref{setup}b inset). In the absence of inertial
effects, this makes no difference to the results. For surface
observations with high-speed video we use a cylindrical cell with
radius $R_{out}=72.25mm$ whose bottom is split at $R_s=55.5mm$.
For the MRI experiments a cell with $R_{out}=43.9mm$ and
$R_s=32.5mm$ is used. The inner disk is rotated at angular
velocities, $\Omega$, in the shear-rate-independent regime. For
the data presented here $\Omega = 1.96rad/s$ and $\Omega =
1.01rad/s$ for the surface and MRI experiments, respectively. We
fill the cell with granular material to a total height $h=H$,
measured from the cell bottom. A layer of grains glued to the cell
walls and bottom assures controlled friction at the boundaries.
For the surface flow measurements we used spherical mustard seeds
($d=1.9mm$) and tracked their motion with high-speed video at
frame rates ranging from 250s$^{-1}$ to 0.027s$^{-1}$. To follow
the motion of particles inside the pack using MRI, we use a
mixture of poppy and rajagara seeds. Rajagara seeds are more
spherical than poppy seeds(Fig.\ref{figure2}b inset), but have
nearly the same average diameter ($d$=0.85mm) and the same density
($\rho$=1.1g/cm$^3$). Poppy seeds contain more oil than rajagara
seeds providing a clear contrast in MRI signal which allows for
particle tracking (Fig.\ref{figure2}a inset).

The simulations are carried out with a discrete element method in
which grains interact only upon contact through assumed point
forces in normal and tangential directions and elastic tangential
displacements are truncated as necessary to satisfy the Coulomb
criteria at the contact. Details of the specific
implementation can be found elsewhere \cite{8}. We
use mono-disperse hertzian spheres with a layer of frozen
particles at the bottom. The relevant parameters describing the
material properties of the spheres are the normal stiffness $k_n=2
\times 10^5mg/d$, the tangential stiffness $k_t=2/7k_n$, the
normal and the tangential viscous damping coefficients $\gamma
_n=50 \sqrt{g/d}$, $\gamma_t=0$, and the particle and wall
coefficients of friction $\mu=0.5$, where $d$ and $m$ are the
diameter and mass of spheres and $g$ is the gravity
acceleration. We have checked that changing the coefficients of
friction does not qualitatively change the observed behavior. The
cell dimensions used for the simulation match those used for the
surface measurement experiments.  We choose $\Omega=0.014\sqrt{d/g}
\simeq 1.39rad/s$.

\begin{figure*}
\begin{center}
\includegraphics[width=6.4in]{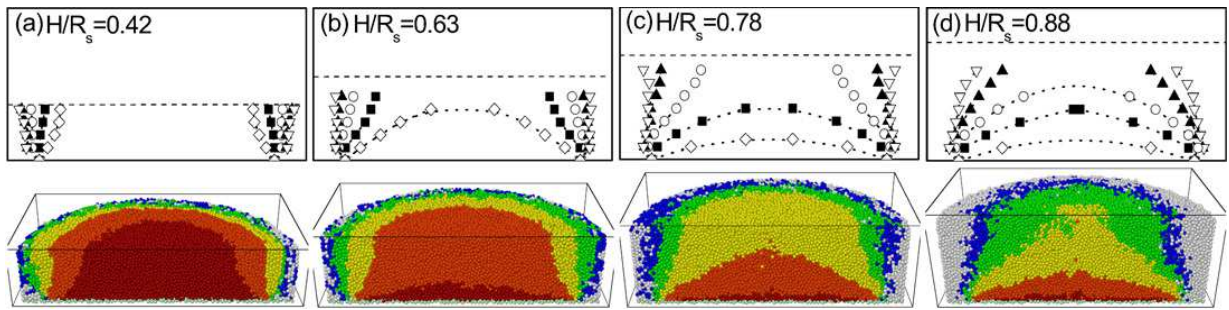}%
\caption{\label{figure3}(color). Contours of constant angular
velocity, $\omega/\Omega$, for different filling height $H$. Upper
panel: MRI experiments. $\omega/\Omega=0.84(\lozenge)$,
$0.24(\blacksquare)$, $2.4 \times 10^{-2}(\circ)$, $2.4 \times
10^{-3}(\blacktriangle)$, $2.4 \times 10^{-4}(\triangledown)$.
Dashed lines indicate $H$ and dotted lines are guides to the eye.
Lower panel: simulations. Color is used to identify velocity
ranges. Dark red: $\omega/\Omega \in [0.84, 1]$; orange: $[0.24,
0.84]$; yellow: $[2.4 \times 10^{-2}, 0.24]$; green: $[2.4 \times
10^{-3}, 2.4 \times 10^{-2}]$; blue: $[2.4 \times 10^{-4}, 2.4
\times 10^{-3}]$; white: $[0, 2.4 \times 10^{-4}]$.}
\end{center}
\end{figure*}

Evidence for a change in flow behavior with increasing filling
height, $H$, can already be found by tracking particle motion at
the free top surface (Fig.\ref{setup}). The angular velocity at
the center of the cell, $\omega_0 = \omega(r=0,h=H)$ is
independent of $H$ for shallow packs but begins to decrease beyond
$H^* \simeq 0.6R_s$. At a similar height Fenistein {\it et al.}
found that the radial shear profile at the surface begins to
deviate from the universal error-function shape describing the
profile for shallow packs \cite{6}. Dramatic deviations are also
seen in the evolution of the center position of the shear band,
$R_c (H)$, with filling height (Fig.\ref{setup}b). Our results for
$R_c$, defined as the radial position where the shear rate has its
maximum, are consistent with previous experiments \cite{6} as well
as with models \cite{7} in the shallow pack regime. However, while
$\omega_0$ decreases for $H>H^*$, the shear zone does not
disappear at the surface as predicted by the theory (dashed line).
Instead, in both experiments and simulations $R_c$ asymptotically
approaches a nonzero value.

These results imply that, beyond $H^*$, velocity gradients must
also exist in the vertical, axial direction near the center of the
cell. Our MRI experiments and simulations explore this shear flow
inside the bulk. We prepare our MRI samples by mixing $5\%$ (by
volume) poppy seeds (MRI positive seeds) uniformly with rajagara
seeds (Fig.\ref{figure2}a inset). Images before rotation and after
an interval of rotation are taken. By performing a
cross-correlation of the two images as a function of radius, we
obtain velocity profiles $\omega(r)$ as a function of $h$
(Fig.\ref{figure2}). This method enables us to measure velocities
with several orders of magnitude difference (for details, see
\cite{16}). For both experiments and simulations care was taken to
assure that the systems are in the steady state by rotating long
enough before any measurement were performed. As an additional
check we made sure that stopping and restarting the system did not
change the velocity profile, which is consistent with the previous
study \cite{11}.

For $H<H^*$, MRI and simulations show an inner core at the center
of the pack which rotates as a solid along with the rotating
bottom. We plot the profiles $\omega(r)$ in Fig.\ref{figure2}a.
There is no slip between different layers near the center of the
cell, and the profiles are fit well by an error function. However,
when we increase $H$ above $H^*$, axial slip occurs in both MRI
experiments and simulations: the bottom layer rotates at the same
rate as the bottom disk, while the layer near the surface hardly
rotates at all (Fig.\ref{figure2}b) \cite{13}. Thus the decrease
in surface flow velocity for $H>H^*$ is caused by shear between
horizontal layers inside the bulk. For $H<H^*$, an inner core
exists at the center of the cell which rotates with the inner disk
with $\frac{\partial \omega (r,h)}{\partial h}\arrowvert_{r=0}=0$;
while for $H>H^*$, $\frac{\partial \omega (r,h)}{\partial
h}\arrowvert_{r=0} \neq 0$.

To visualize the resulting shear profiles, we plot cross-sections
of the system with contours of constant angular velocity
(Fig.\ref{figure3}). For $H$ well above $H^*$, the high-velocity
contours close into dome shapes (Fig.\ref{figure3}d), which
gradually open as $H$ decreases. The contours for smaller
velocities open up earlier and eventually touch the surface
(Fig.\ref{figure3}b, c). When $H<H^*$, all velocity contours touch
the surface as a solid inner core forms for $r < R_s$
(Fig.\ref{figure3}a).  The lower panel shows the corresponding
simulation results in a color gradient. One can see how the motion
at different $h$ correlates with the motion of the bottom disk.

\begin{figure}
\begin{center}
\includegraphics[width=3.0in]{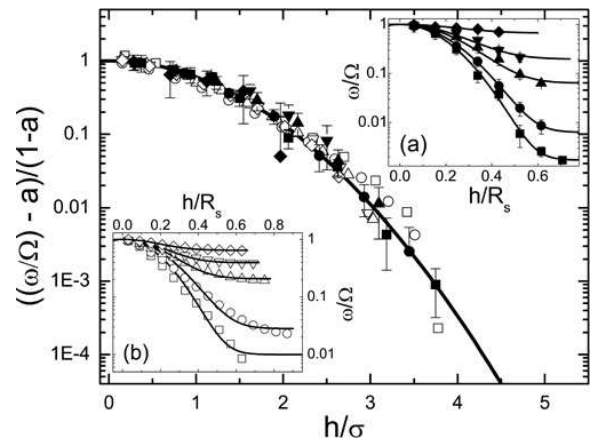}%
\caption{\label{figure4} Scaled angular velocity
$((\omega/\Omega)-a)/(1-a)$ for different heights $h/\sigma$.
Insets show the corresponding unscaled data from (a) MRI and (b)
simulations. The main panel uses the same symbols as the insets.
Solid line is a gaussian exp$(-h^2/(2\sigma^2))$ with
$\sigma=0.18R_s$. Inset (a): $H=0.97R_s (\blacksquare)$,
$H=0.88R_s (\bullet)$, $H=0.78R_s (\blacktriangle)$, $H=0.74R_s
(\blacktriangledown)$ and $H=0.63R_s (\blacklozenge)$. Inset (b):
$H=1.02R_s (\square)$, $H=0.90R_s (\circ)$, $H=0.78R_s
(\triangle)$, $H=0.72R_s (\triangledown)$ and $H=0.66R_s
(\lozenge)$. Solid lines are gaussian fits introduced in the
text.}
\end{center}
\end{figure}

\begin{figure}
\begin{center}
\includegraphics[width=3.0in]{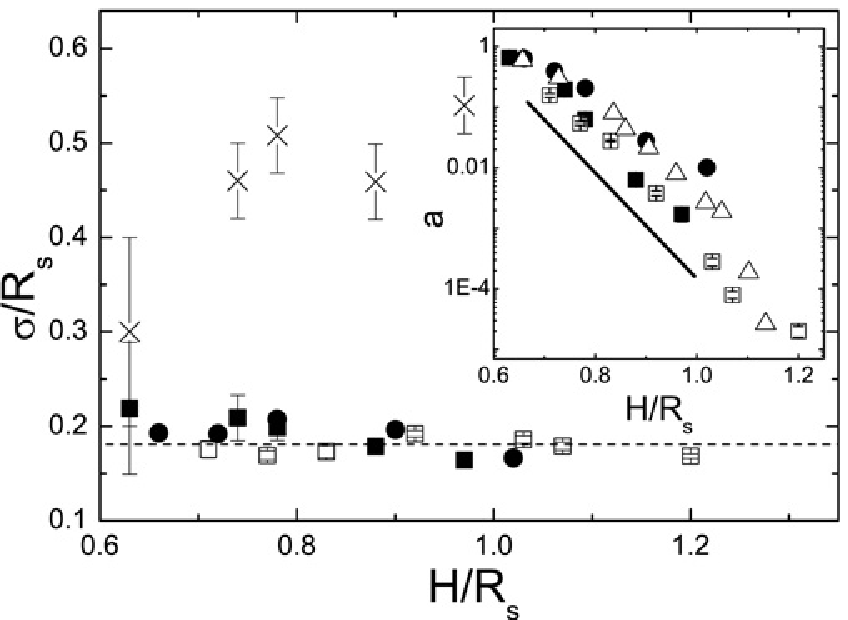}%
\caption{\label{figure5} Width of the shear profiles, $\sigma$,
for different filling heights $H$. Data show $\sigma$ both along
the axial direction in the bulk ($\blacksquare$ and $\square$ from
MRI experiments using different methods \cite{12}, $\bullet$ from
simulations) and along the radial direction in the surface layer
($\times$). Dashed line is $\sigma=0.18R_s$. Inset: offset $a$ vs.
$H$ from MRI ($\blacksquare$, $\square$) and simulations
($\bullet$). Surface data from Fig.\ref{setup} are shown for
comparison ($\triangle$). Solid line indicates exponential
behavior.}
\end{center}
\end{figure}

Having access to the full shear profiles allows us to address the
question whether the same length scale describes the shear in
axial and radial directions. Fig.\ref{figure4} inset a, b show
$\omega (h)$ measured along the axis of the cell ($r=0$) for
different heights $H$. We note that the axial velocity decays in
an exponential fashion before the data plateau at a level that
extends to the surface. All MRI and simulation data can be fit
consistently to a gaussian form,
$\omega(h)/\Omega=a+(1-a)$exp$[-h^2/(2\sigma^2)]$, where $a$ is an
$H$-dependent offset indicating the angular velocity at the top
surface and $\sigma$ is the axial shear-band width. The main panel
in Fig.\ref{figure4} demonstrates the consistency of fitting a
gaussian to the axial shear profile, collapsing data from MRI
measurements and simulations. Near $H^*$, the offset, $a$,
approaches unity in an exponential manner (Fig.\ref{figure5}
inset). Although all data show the same exponential behavior, the
offsets from simulations and surface measurement with
mono-disperse spherical grains are larger than those from the MRI
experiments with poppy and rajagara seeds \cite{13}. Extrapolating
each set of data to $\omega/\Omega=1$ suggests that the onset of
axial shear begins at $H^*=0.60\pm 0.02R_s$.

The data in Fig.\ref{figure5}, together with those from
Fig.\ref{setup}a, demonstrate that the transition in shearing
behavior at $H^*$ occurs in a continuous manner. This differs from
the model by Unger {\it et al.} \cite{7} which predicts a
first-order transition at $H^*=0.7R_s$. While the Unger model,
based on the idea of minimum dissipation of energy, includes the
essential elements for the transition in the shape of the shear
band, a key aspect not considered is the axial slip below $H^*$.
The necessity for such slip emerges from considering the torque
balance in shallow packs. When the torque on the surface of the
inner core exceeds the frictional strength at the bottom, slip
will occur. The total torque on the surface of the inner core and
the torsional strength of the contact between the bottom of the
cell and the granular material above it can be calculated if the
shape of the inner core $h_{core}(r)$ is known. Using the
approximate $h_{core}(r)$ given in \cite{7}, it can be shown that
this already occurs when $H > 0.5R_s$.  Thus, already within the
Unger model a torsional failure mode near the bottom should
preempt any first-order transition within the bulk. Our data
demonstrate that this torsional failure is associated with a
well-defined axial shear band at $r=0$ that exhibits a gaussian
profile.

Gaussian shear profiles have been observed in previous studies
using traditional Couette cells \cite{4}. However, in the present
geometry we find such a profile along the axial direction and an
approximate error-function profile along the radial direction. The
width of this axial shear zone appears to be independent of $H$,
with an average value $\sigma=0.18R_s$, implying $\sigma/d=6.9$
for the experiment and $5.4$ for the simulation
(Fig.\ref{figure5}). This is in contrast to the width of the
radial shear profile (Fig.\ref{figure5}), which strongly depends
on $H$ and approaches 0 as $H\rightarrow0$ \cite{6}.

The ubiquitous presence of shear bands is one of the crucial
differences between granular materials and ordinary fluids.
Understanding what gives rise to the shear profiles is one of the
outstanding puzzles in granular dynamics.  The modified Couette
geometry produces two distinct forms of shear: a radial component
whose width grows with height as shown by Fenistein {\it et al.}
\cite{6}, and an axial component with a small constant width that
only appears when the filling height exceeds a threshold. The
different character of the shear bands in the radial and axial
directions shows that boundary conditions are essential for
determining shear localization. Because the onset of axial shear
in this geometry is continuous and controlled simply by the height
of the pack, these studies have allowed for detailed observation
of how shear can be initiated in the bulk. Very recent surface
measurement of the central core procession \cite{15} also
corroborates our three dimensional results.

\begin{acknowledgments}
We acknowledge E. Corwin, X. Fan, D. Fenistein, M. van Hecke and
J. River. Collaboration was performed under the auspices of DOE
Center of Excellence for the Synthesis and Processing of Advanced
Materials. Sandia is a multiprogram laboratory operated by Sandia
Corporation, a Lockheed Martin Company, for the US DOE National
Nuclear Security Administration under contract DE-AC04-94AL85000.
Work at the UofC was supported by NSF MRSEC DMR-0213745, NSF
CTS-0405619, DOE DE-FG02-03ER46088 and MAT2003-03051-C03-01
(Spanish Government). AFB thanks the Spanish Ministerio de
Educaci{\'o}n y Ciencia for a grant to enable an extended stay at
the UofC.

\end{acknowledgments}


\begin{thebibliography}{99}
\bibitem{13} We believe that the quantitative difference between MRI experiments and
simulations is due to the different cell geometry and particle
shape used in the two situations. However, they all show the same
trend.
\bibitem{12} ($\blacksquare$) is measured with the method introduced in the text.
($\square$) is obtained by placing a thin cross of poppy seeds at
the center of the cell and measuring its distortion after
rotation, smiliar to \cite{6}. For a detailed discussion of the
experimental methods, see \cite{16}.
\end{thebibliography}
\end{document}